\documentclass[reprint, amsmath,amssymb, aps, prb, superscriptaddress]{revtex4-1}
\pdfoutput=1

\usepackage{geometry}
\usepackage{mathrsfs}
\usepackage{amsmath}
\usepackage{graphics}
\usepackage{color}
\usepackage{courier}
\usepackage{url}
\usepackage{listings}
\usepackage[table,xcdraw]{xcolor}
\usepackage[]{subfig}
\usepackage{floatrow}
\usepackage[colorlinks]{hyperref}
\hypersetup{breaklinks=true}
\usepackage{cleveref}
\usepackage{float}
\floatstyle{plaintop}
\usepackage{adjustbox}
\restylefloat{table}
\usepackage{xcolor}
\usepackage{ulem}

\newcommand{\gts}{\text{GaTa}_4\text{Se}_8}

\newcommand{\gns}{\text{GaNb}_{4}\text{S}_8}
\newcommand{\gnse}{\text{GaNb}_{4}\text{Se}_8}
\newcommand{\fcc}{F\bar{4}3m}
\newcommand{\rh}{R3m}
\newcommand{\Pmn}{P\bar{4}2_1m}
\newcommand{\trm}{P\bar{4}m2}

\begin{document}
\preprint{APS/123-QED}

\title{Crystal and electronic structure of $\gts$ from first-principles calculations}

\author{Shuai Zhang}
\affiliation{Beijing National Laboratory for Condensed Matter Physics, and Institute of Physics, Chinese Academy of Sciences, Beijing 100190, China}
\affiliation{University of Chinese Academy of Sciences, Beijing 100049, China}

\author{Tiantian Zhang}
\affiliation{Beijing National Laboratory for Condensed Matter Physics, and Institute of Physics, Chinese Academy of Sciences, Beijing 100190, China}
\affiliation{Department of Physics, Tokyo Institute of Technology, Ookayama, Meguro-ku, Tokyo 152-8551, Japan} 
\affiliation{Tokodai Institute for Element Strategy, Tokyo Institute of Technology,Nagatsuta, Midori-ku, Yokohama, Kanagawa 226-8503, Japan} 

\author{Hongshan Deng}
\affiliation{Center for High Pressure Science and Technology Advanced Research, Beijing 100094, China}

\author{Yang Ding} 
\email{yang.ding@hpstar.ac.cn} 
\affiliation{Center for High Pressure Science and Technology Advanced Research, Beijing 100094, China}

\author{Yue Chen}
\email{yuechen@hku.hk}
\affiliation{Department of Mechanical Engineering, The University of Hong Kong, Pokfulam Road, Hong Kong SAR, China}

\author{Hongming Weng}
\email{hmweng@iphy.ac.cn}
\affiliation{Beijing National Laboratory for Condensed Matter Physics, and Institute of Physics, Chinese Academy of Sciences, Beijing 100190, China}
\affiliation{University of Chinese Academy of Sciences, Beijing 100049, China}
\affiliation{Songshan Lake Materials Laboratory, Dongguan, Guangdong 523808, China}

\begin{abstract}
    GaTa$_4$Se$_8$ belongs to the lacunar spinel family and there have been intensive studies on its novel properties, 
    such as its possible Mott-insulator state and  superconductivity under pressure. However, its crystal structure and phase transition are still not well known. 
    In this work, we investigated three different crystal structures, proposed in experiments,  using first-principle calculations. 
    For the cubic phase with space group $F\bar{4}3m$, its phonon spectra have three soft modes in the whole Brillouin zone, indicating the strong dynamical instability. 
    The second one is the trigonal phase with space group $R3m$, which has been proposed based on Raman spectra under high pressure. 
    This phase can be deduced from the soft phonon modes at $\Gamma$ of the cubic phase and it is dynamically stable according to its phonon spectra. 
    The third one is the tetragonal phase with space group $\Pmn$, which is also stable according to its phonon spectra and may be the low-temperature phase from x-ray diffraction.
    Within local density approximation calculations, the  cubic and trigonal phases are metals, while the tetragonal phase is 
  a band insulator consistent with the insulating feature in experiments.  Our results suggest the possibility of the non-Mott state of GaTa$_4$Se$_8$ at low temperature and ambient pressure as a result of lattice distortion. 
    On the other hand, the electronic structure of the trigonal phase can be viewed as a single-band Hubbard model. 
    The Mott insulator state has been obtained within dynamical mean field theory calculation when the interaction parameter $U$ is larger than 0.40 eV vs. a bandwidth of 0.25 eV. 
    We hope these findings would be helpful in solving the long-standing problem of the ambiguity in the structural phase of GaTa$_4$Se$_8$.
\end{abstract}

\maketitle

\section*{Introduction}
$\gts$ belongs to the lacunar spinel compounds family with $AM_4X_8$ as their chemical formula unit~\cite{camjayi_first-order_2014}, 
  where $A$=Ga or Ge, $M$=V, Mo, Nb  or Ta, and $X$=S or Se. It usually takes the 
  typical cubic structure with a space-group symmetry of $F\bar{4}3m$~\cite{yaich_gts_crystal_1984}. The $M$ atoms can be  considered to form interconnected $M_4$ clusters and they are thought to be responsible for most intriguing properties of the family compounds. $\gts$, as well as $\gns$ and $\gnse$, has been proposed to be a Mott insulator at room temperature
  and ambient pressure~\cite{pocha_crystal_2005}. $\gts$ has no long-range magnetic order when the temperature is down to 1.6 K~\cite{abd-elmeguid_transition_2004}. Under pressure, it has an insulator/metal coexistent state and a hysteresis phenomenon in resistivity vs. temperature~\cite{camjayi_first-order_2014}. It has also been reported to have pressure-induced superconductivity at low temperatures~\cite{abd-elmeguid_transition_2004}. 
  These have been ascribed to the electrons from the molecular orbitals of Ta$_4$ clusters since these electrons  have  strong spin-orbit coupling (SOC) and they are localized on the clusters with reduced kinetic energy due to the long-distance separation of the clusters.
  These make $\gts$ an ideal platform to explore the correlation physics among the electrons with non-negligible SOC forming $j_{eff}$ molecular orbitals on the $M_4$ clusters~\cite{kim_spin-orbital_2014, jeong_direct_2017, jeong2020novel_superconduct_topo, park2020pressure_topo_superconductor}.  
 
   In addition to the correlation effects among Ta$_4$ molecular orbitals,  the changes in crystal structure
   can affect the electronic structures of these compounds substantially~\cite{sieberer_importance_2007}, and the above novel physical properties may also be related to the changes.
 The tetragonal $\Pmn$ structure was proposed as the low-temperature and ambient-pressure phase in 2007~\cite{jakob_phd_2007}, and the trigonal
 $\rh$ structure was proposed as a possible high-pressure phase in 2009~\cite{muller_phd_2007}.
    However, very recent experiments proposed that $\gts$  may have a structural phase transition to another tetragonal phase of $\trm$ according to the powder x-ray diffraction (XRD) pattern at low temperature~\cite{japan_gts} and a high-pressure monoclinic phase of space group $C2$ from XRD  and Raman spectra~\cite{Ding_gts_2020}.
   These disputes  motivate us to study the crystal structures of $\gts$ in detail based on first-principles calculations. 
   In fact, there have been several first-principles calculations  on $AM_4X_8$ family compounds~\cite{sieberer_importance_2007, dft_functionals_lunar_spninel,kim_spin-orbital_2014,jeong_direct_2017}. For $\gts$ with the above cubic and trigonal phases, some of these calculations showed that a band gap at the Fermi level can be achieved when a kind of magnetic ordering is artificially assumed~\cite{dft_functionals_lunar_spninel},
 which is inconsistent with its nonmagnetic ground-state feature in experiments~\cite{japan_gts}.
 
 In this work, we  performed first-principles calculations to study the phonon spectra and electronic structure of $\gts$ for its three different crystal structures proposed in experiments, namely, the cubic ($\fcc$), trigonal ($\rh$), and tetragonal ($\Pmn$) phases.
 The other tetragonal ($\trm$) and monoclinic ($C2$)  phases  were not studied since the complete crystal structural information  is lacking.
 The soft modes at $\Gamma$ in the $\fcc$ phase can lead to the $\rh$ phase, of which the phonon spectra have no image frequency. The nonmagnetic first-principles calculations show these two phases are metallic, and the insulating state of these two phases can be obtained when correlation effects among the Ta$_4$ molecular orbitals are considered in Ref.~[\onlinecite{camjayi_first-order_2014}] (for the $\fcc$ phase) and this work (for the $\rh$ phase).
 For the $\Pmn$ phase, its phonon spectra also have no imaginary frequency. Our calculations show it can be an ordinary band insulator due to the tetramerization of Ta$_4$ clusters from the cubic phase.

  This paper is organized as follows: We first introduce the calculation method and then discuss the three different crystal structures 
in the symmetry lowering order. In each case, the phonon spectra and electronic structure are shown and discussed. Finally, further discussions are made.

\section{Methodology}
We used the Vienna \textit{Ab initio} Simulation package  (\texttt{VASP})~\cite{kresse1996efficient, perdew_generalized_1996}
  and the \texttt{PHONOPY} package~\cite{phonopy} within the density-functional perturbation theory (DFPT)~\cite{dfpt_rev} scheme 
  to perform the phonon spectra calculation. 
    In the phonon spectra calculation  of $\gts$,  for both $\fcc$ and $\rh$ space groups, we set the Monkhorst-Pack $k$-point mesh of 3 $\times$ 3 $\times$ 3 and plane-wave cutoff energy of 420 eV.
    For the structure of $\Pmn$, we used the $k$-point mesh of $2 \times 2 \times 2$ and the same plane-wave energy cutoff.
    We used a $2 \times 2 \times 2$ supercell of the primitive  cell in the phonon spectra calculation for all structures.
    In all calculations, the projector-augmented-wave (PAW) method~\cite{blochl1994projector,kresse1999ultrasoft} with the Perdew-Burke-Ernzerhof (PBE) 
  exchange–correlation functional~\cite{perdew_generalized_1996} was used.
    For electronic structure calculation, the plane-wave energy cutoff  was also set as 420 eV with a $7 \times 7 \times 7$ Monkhorst-Pack $k$-point mesh in the self-consistent-field calculation. 
  Phonon spectra calculations were performed without considering the SOC, and all calculations were performed with the nonmagnetic state because of the absence of  long-range magnetic order.

\section{Results and Discussions}
    \subsection{$\fcc$ structure}
\begin{figure}
      \centering
      \sidesubfloat[]{
        \includegraphics[width=0.45\linewidth]{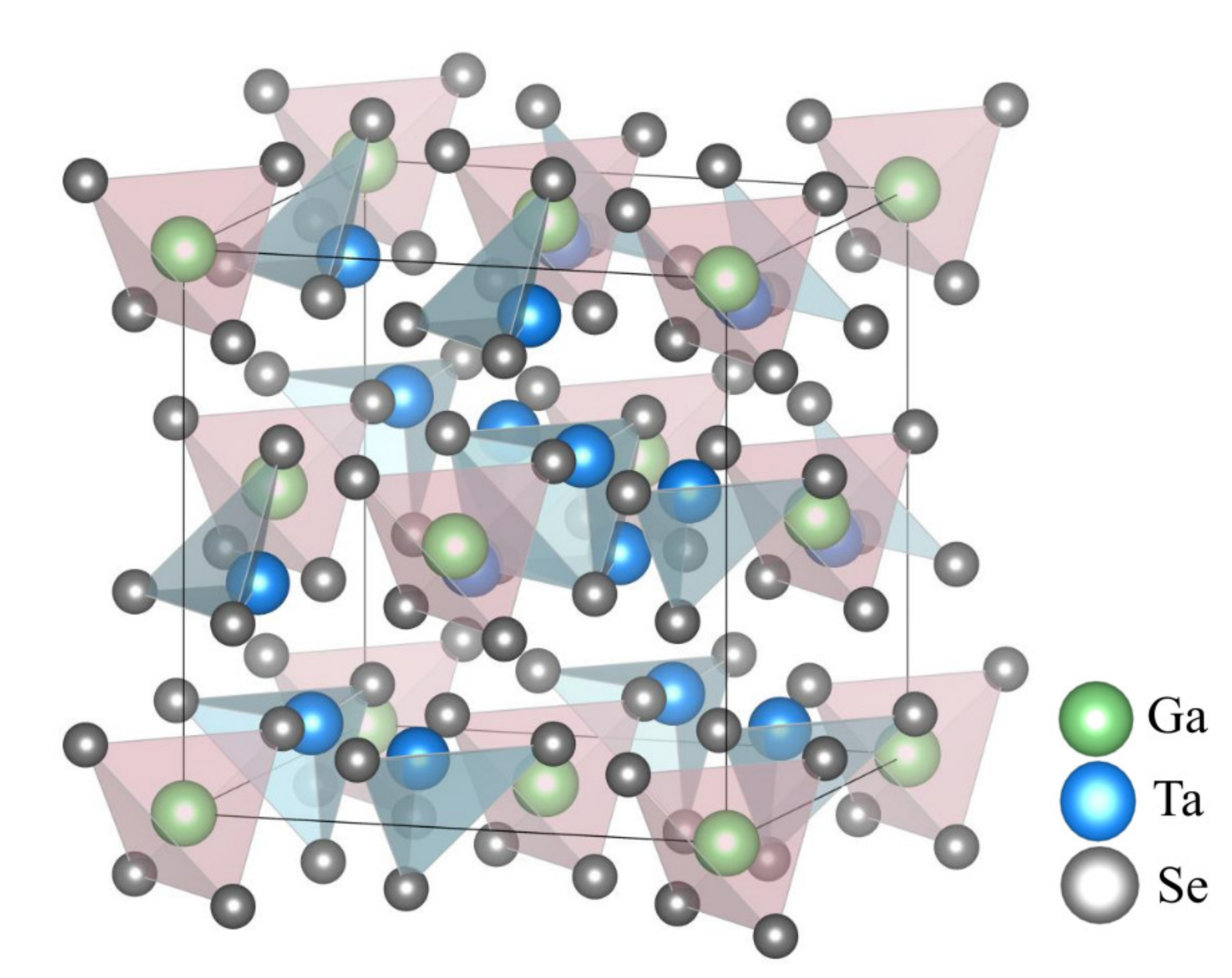}
        \label{fig:crys216}
      }
      \sidesubfloat[]{
        \includegraphics[width=0.3\linewidth]{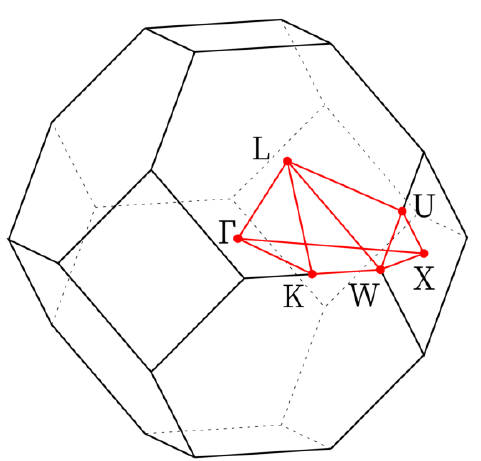}
        \label{fig:BZ216}
      }
    
      \sidesubfloat[]{
        \includegraphics[width=0.9\linewidth]{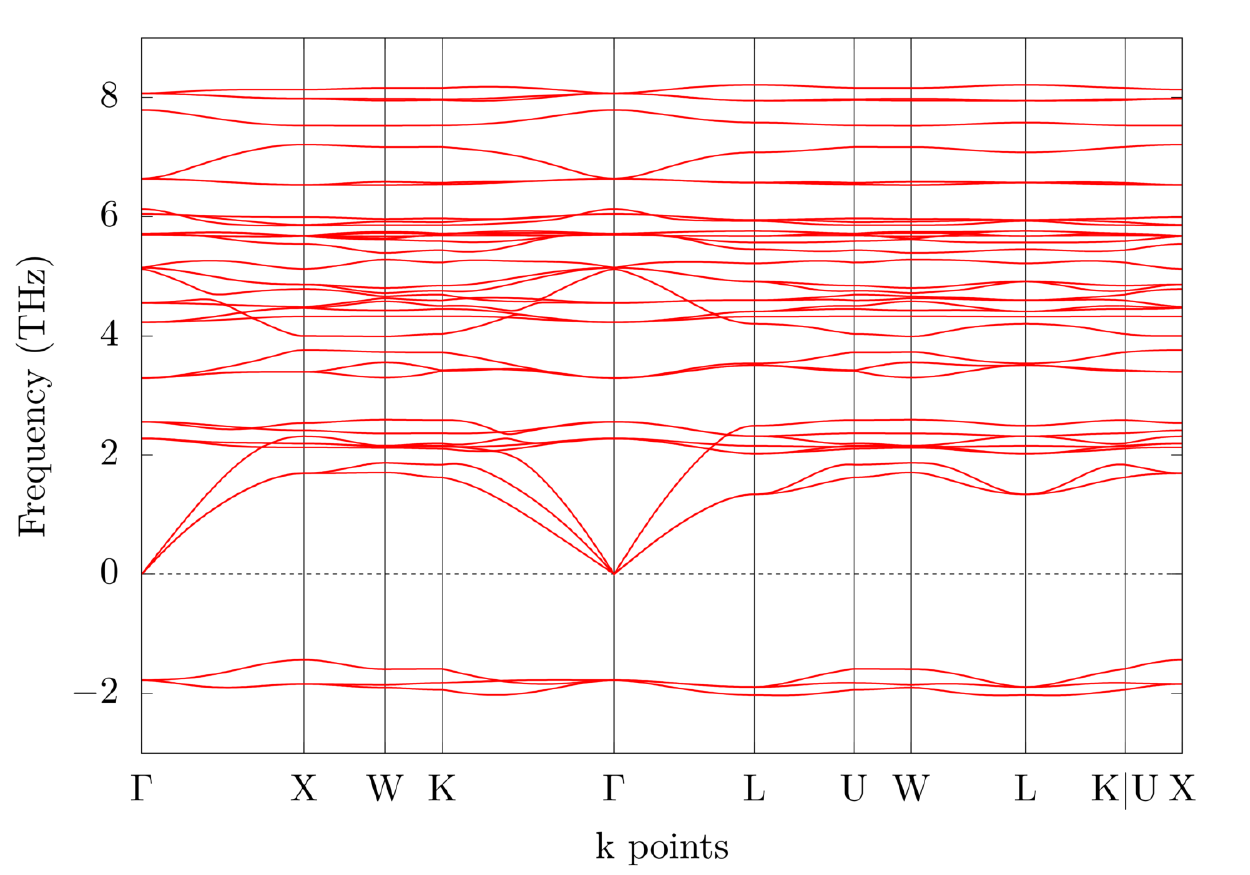}
        \label{fig:phonon216_f}
      }
      \caption{
        \ref{fig:crys216} Crystal structure of $\fcc$  $\gts$, 
        \ref{fig:BZ216} its Brillouin zone, and 
        \ref{fig:phonon216_f} the phonon spectra. 
      }
      \label{fig:crys_bz216}
    \end{figure}

    The lacunar spinel $AM_4X_8$ family has a fcc Bravais lattice with space group $\fcc$ (No. 216). The conventional unit cell of $\gts$ and its first Brillouin zone are shown in \Cref{fig:crys216} and \Cref{fig:BZ216}, respectively. 
    It can be derived from the spinel structure $AM_2X_4$ of space group $Fd\bar{3}m$~\cite{pocha_crystal_2005, gav4s8_r3m}.
    Ga (atom $A$) occupies one-half of the tetrahedral sites  in the cubic close packing of Se (atom $X$) atoms in an ordered way. Ta (atom $M$) shifts along the $C_{3v}$ axis ($u$,$u$,$u$) from $u=0.625$ to $u\approx0.602$. 
    This shift of Ta leads to the formation of Ta$_4$ tetrahedral clusters with intracluster $M$-$M$ distance of 3.001 \AA ~and intercluster distance of 4.339 \AA (the experimental structure at room temperature~\cite{springer_gts_crys}). Thus, the structure can be appropriately considered as 
  a rocksaltlike arrangement of Ta$_4$Se$_4$ cubes and GaSe$_4$ tetrahedrons~\cite{pocha_crystal_2005}.
    However, as shown in \Cref{fig:phonon216_f}, the phonon spectra of this $\fcc$ structure  have large imaginary frequencies in the whole Brillouin zone, 
  indicating the dynamical instability of this structure. At $\Gamma$ , the three soft modes are degenerate, which is consistent with the cubic symmetry 
  that the possible spontaneous symmetry breaking is equivalent in three directions.

\begin{figure}
      \centering
      \sidesubfloat[]{
        \includegraphics[width=0.9\columnwidth]{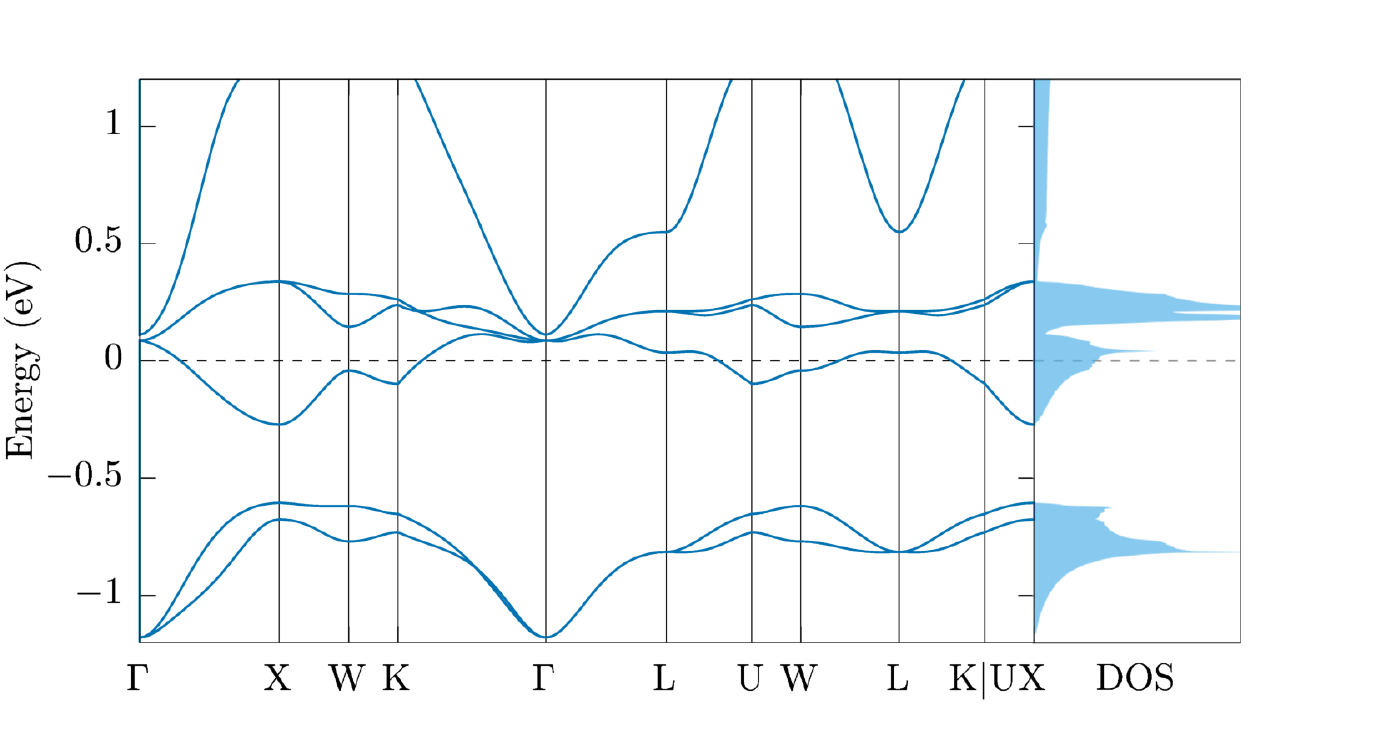}
        \label{fig:band216_nosoc}
      }

      \sidesubfloat[]{
        \includegraphics[width=0.9\columnwidth]{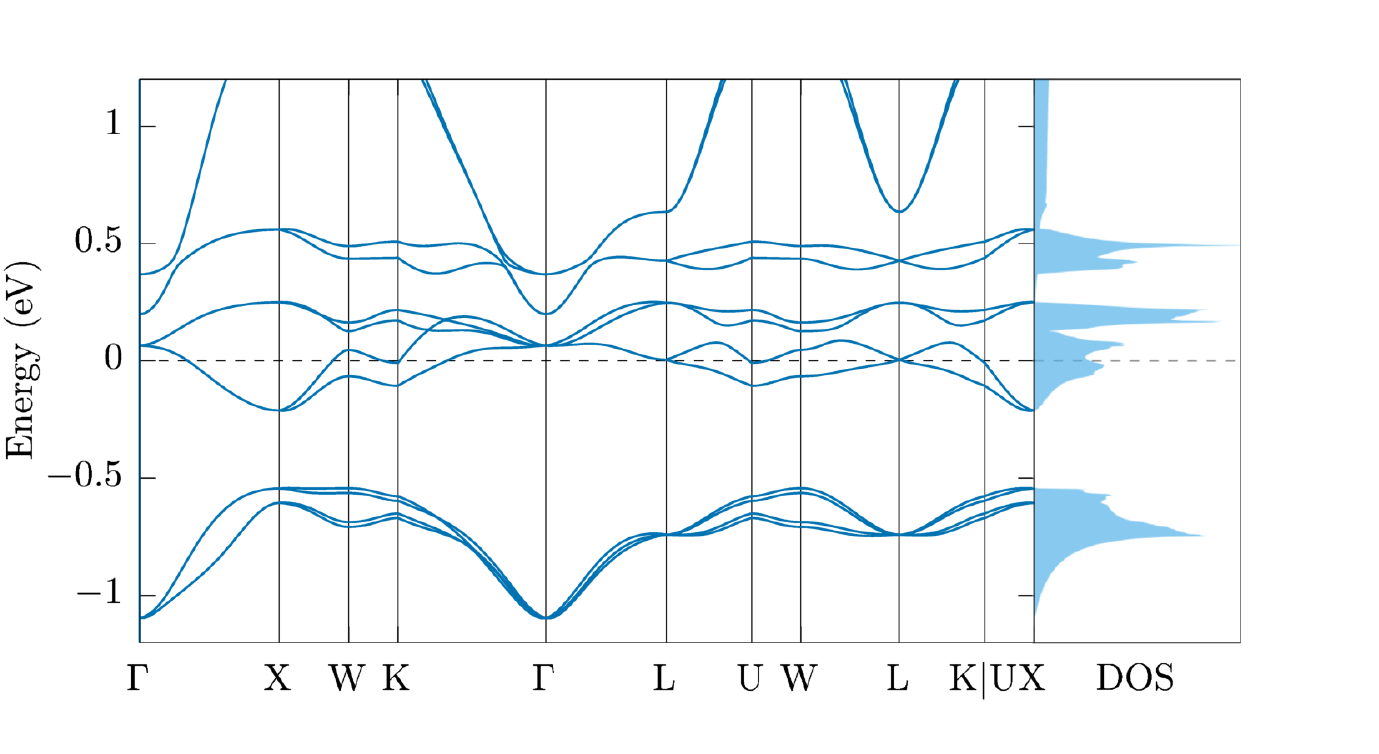}
        \label{fig:band216_soc}
      }
      \caption{The band structure and density of states (DOS) for nonmagnetic $\gts$ of $\fcc$ structure \ref{fig:band216_nosoc} without and \ref{fig:band216_soc} with SOC.
      }
      \label{fig:band216} 
    \end{figure}

    The electronic band structures in \Cref{fig:band216} show no gap at the Fermi level in calculations with and without spin-orbit coupling. This indicates that $\fcc$  $\gts$ is a metal in our single-particle approximation calculation. The  narrow bands around the Fermi level are  composed of molecular orbitals from Ta$_4$ clusters. Based on this, the Mott insulator state for  $\fcc$  $\gts$ has been obtained in dynamical mean-field theory  (DMFT) calculation~\cite{camjayi_first-order_2014} when the correlation effects among these molecular orbitals are considered. As it was mentioned before, there is evidence showing that $\gts$ may have structural phase transitions while the temperature decreases, and the structural distortion may change the electronic states dramatically. Therefore, it is necessary to investigate  the possibility of the ordinary band-insulator state resulting from the lattice distortion, instead of taking it as a Mott insulator directly at low temperatures.

\begin{figure}
      \centering
      \includegraphics[width=\linewidth]{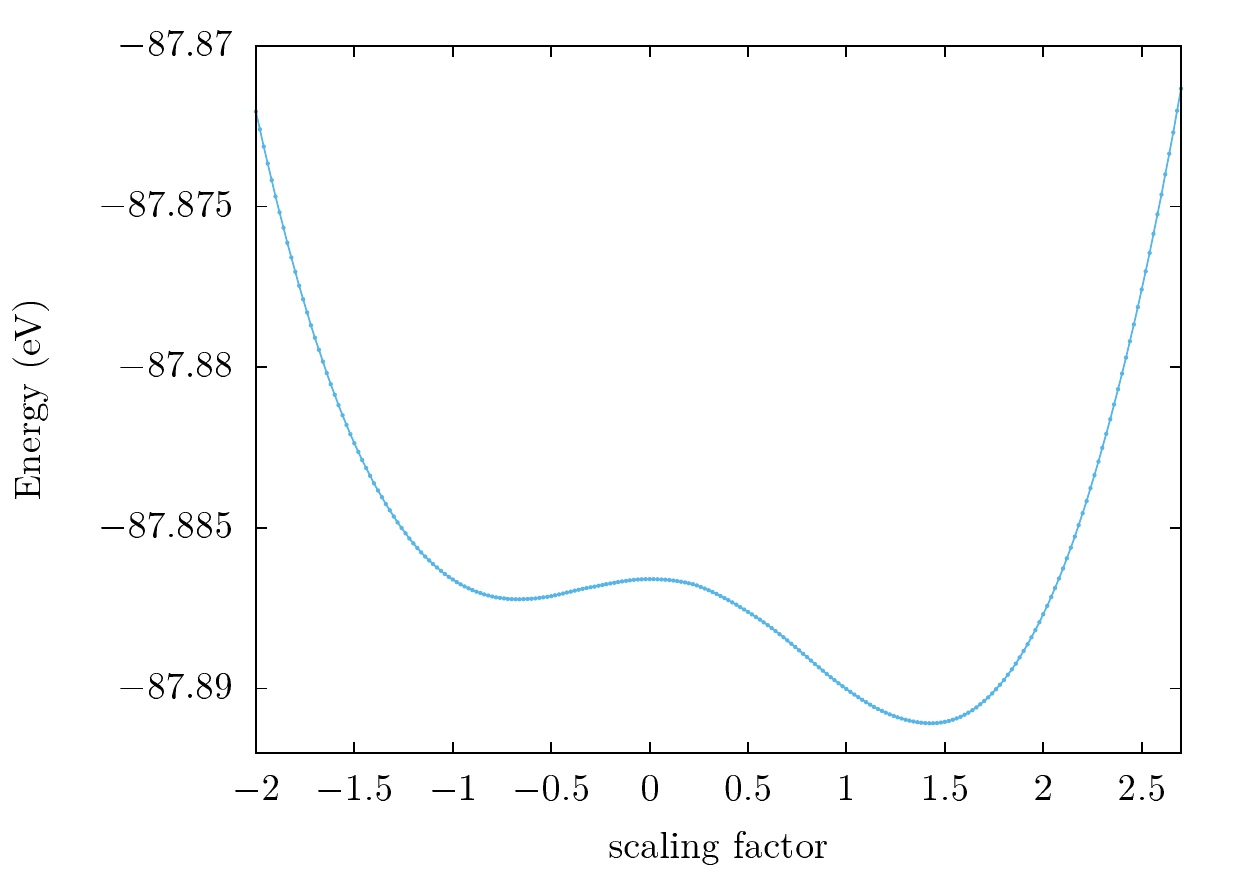}
      \caption{\label{fig:well} The asymmetric ``double" well of the $\fcc$ structure in plot of total energy vs distortion originated from the absence of inversion symmetry. The scaling factor is 
      a dimensionless number to indicate the size of distortion away from the $\fcc$ structure.}
    \end{figure}

    The soft-mode analysis is a common method to capture possible phase transitions within first-principles calculations.  It is known that a complete investigation of all the linear combinations of  soft modes is  infeasible. Herein, we  take one of the three degenerate soft modes at $\Gamma$ for the analysis to explore how and what structural phase transition will happen if it is frozen.  
    We multiply the mode with a dimensionless scaling factor varying from -2 to 3 in 250 steps to obtain various atomic displacements away from the original primitive cell of the $\fcc$ structure. 
    After the total energy self-consistent calculations for each crystal structure with the above structural distortions, we get an asymmetric double well of total energy vs. distortion, as shown in \Cref{fig:well}. It is noted that the original cubic structure is noncentrosymmetric and the distortions with the above positive scaling factor are not symmetric with that of the opposite sign. This is different from the symmetric double well in the transition from  paraelectric  to ferroelectric  phase. The structure with the scaling factor around 1.5 has the lowest total energy. We found that this distorted structure has a space group $\rh$ (No. 160). 
    In fact, the $\rh$ phase is commonly  seen in lacunar spinel materials, such as GaV$_4$Se$_8$~\cite{ganb4se8_r3m_phd}, GaV$_4$S$_8$~\cite{gav4s8_r3m}, and GaMo$_4$S$_8$~\cite{gamo4s8_r3m}. 
    There was already a proposal that the $\rh$ structure may be a high-pressure phase of $\gts$ according to the Raman spectra~\cite{muller_phd_2007}.
    We further investigate this phase in the next section. 

  \subsection{$\rh$ structure}
\begin{figure}
      \centering
      \sidesubfloat[]{
        \includegraphics[width=0.45\linewidth]{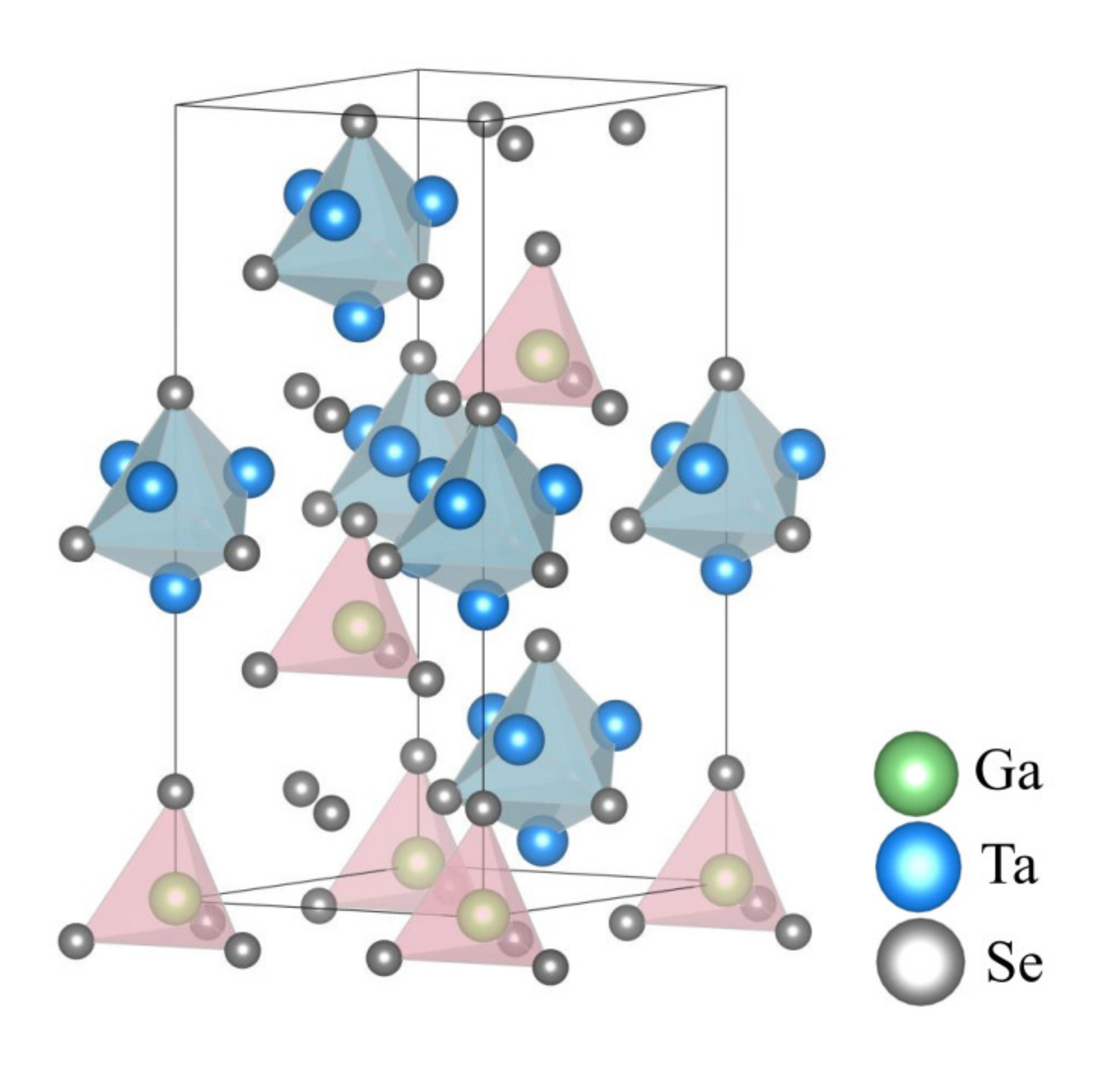}
        \label{fig:crys160}
      }
      \sidesubfloat[]{
        \includegraphics[width=0.35\linewidth]{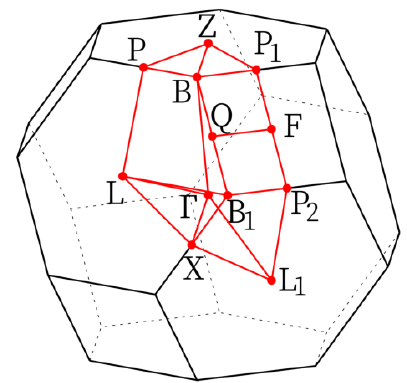}
        \label{fig:BZ160}
      }

      \sidesubfloat[]{
        \includegraphics[width=0.9\linewidth]{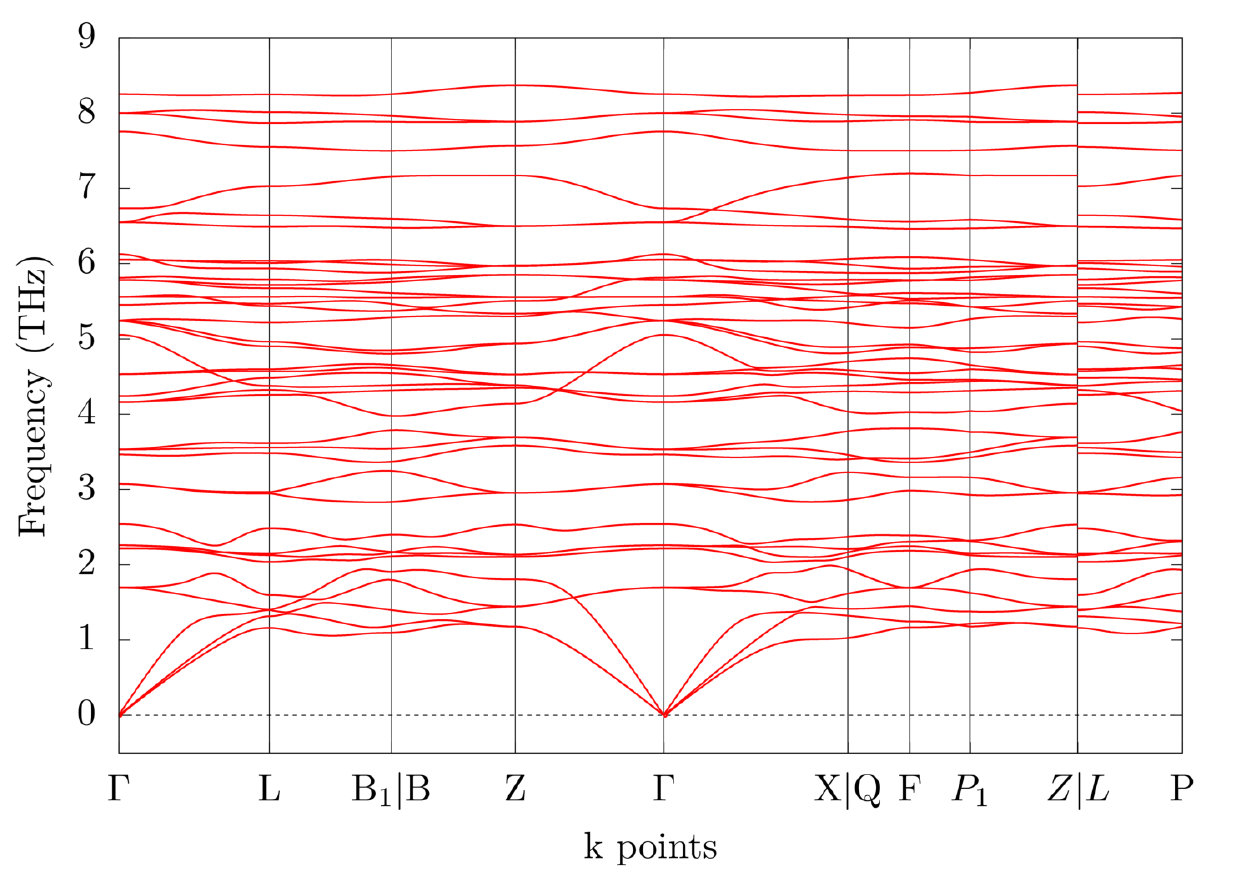}
        \label{fig:phonon160}
      }
      \caption{
        \ref{fig:crys160} Crystal structure, \ref{fig:BZ160} Brillouin zone,  and \ref{fig:phonon160} the phonon spectra of $\rh$ $\gts$. 
      }
    \end{figure}

      The conventional unit cell of $\rh$ $\gts$ is shown in \Cref{fig:crys160},
    which is obtained through the full structural optimization of the distorted crystal
    structure at the bottom of the lower potential well in \Cref{fig:well}.
    In the $\rh$ phase, the Ta$_4$ clusters are elongated along  the $C_{3v}$
    axis. As a result, the original regular tetrahedron Ta$_4$ clusters are transformed into regular triangular pyramids, with two different bond lengths: 2.950 \AA (base face bond length) and 3.119 \AA  (lateral edge bond length).
      There is  one chemical formula unit in the primitive  cell and the volume of the primitive cell is  slightly larger
    than that of the $\fcc$ structure. The calculated total energy per chemical formula for the fully optimized trigonal $\rh$ phase is lower than that of $\fcc$ by 0.01 eV.
      The phonon spectra in \Cref{fig:phonon160}  show  no imaginary frequency in the whole Brillouin zone, indicating that  $\rh$ $\gts$ is dynamically stable.

    \begin{figure}
      \centering
      \sidesubfloat[]{
        \includegraphics[width=0.9\columnwidth]{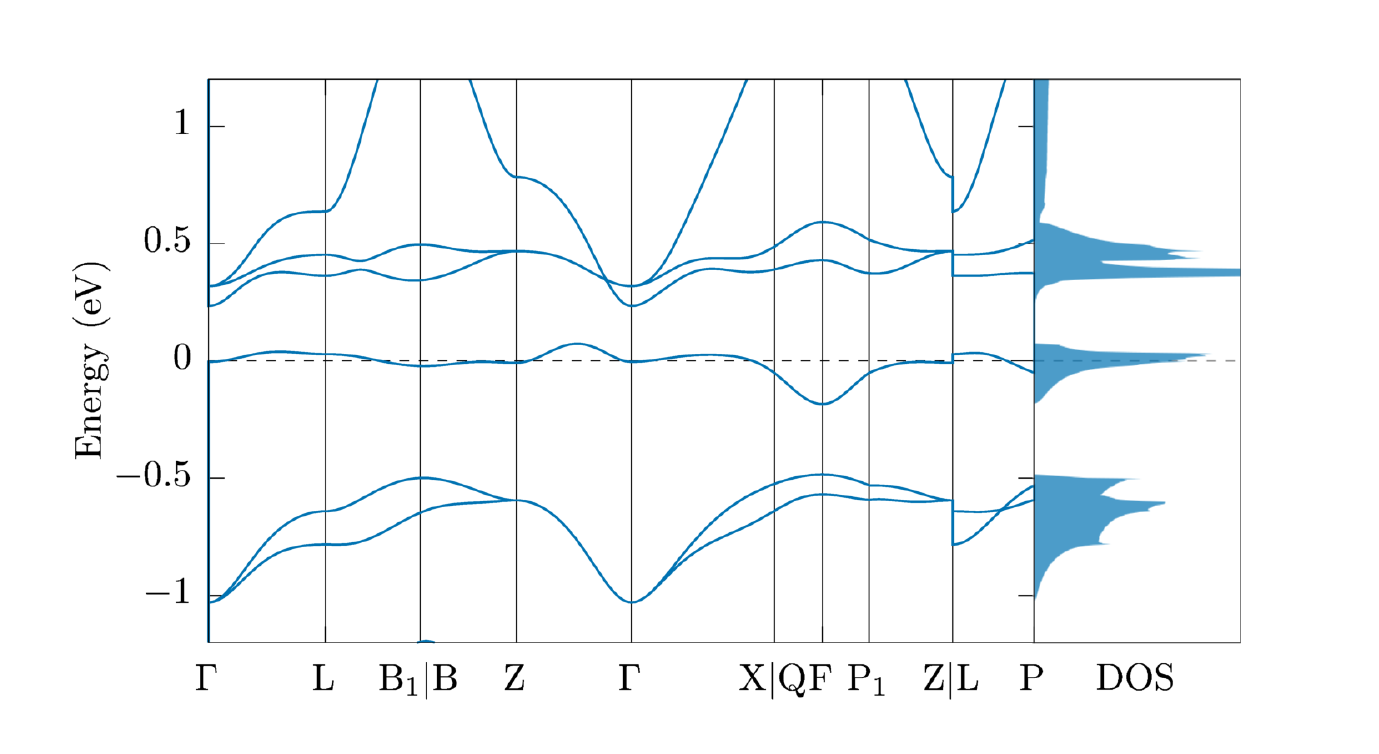}
        \label{fig:band160_nosoc}
      }

      \sidesubfloat[]{
        \includegraphics[width=0.9\columnwidth]{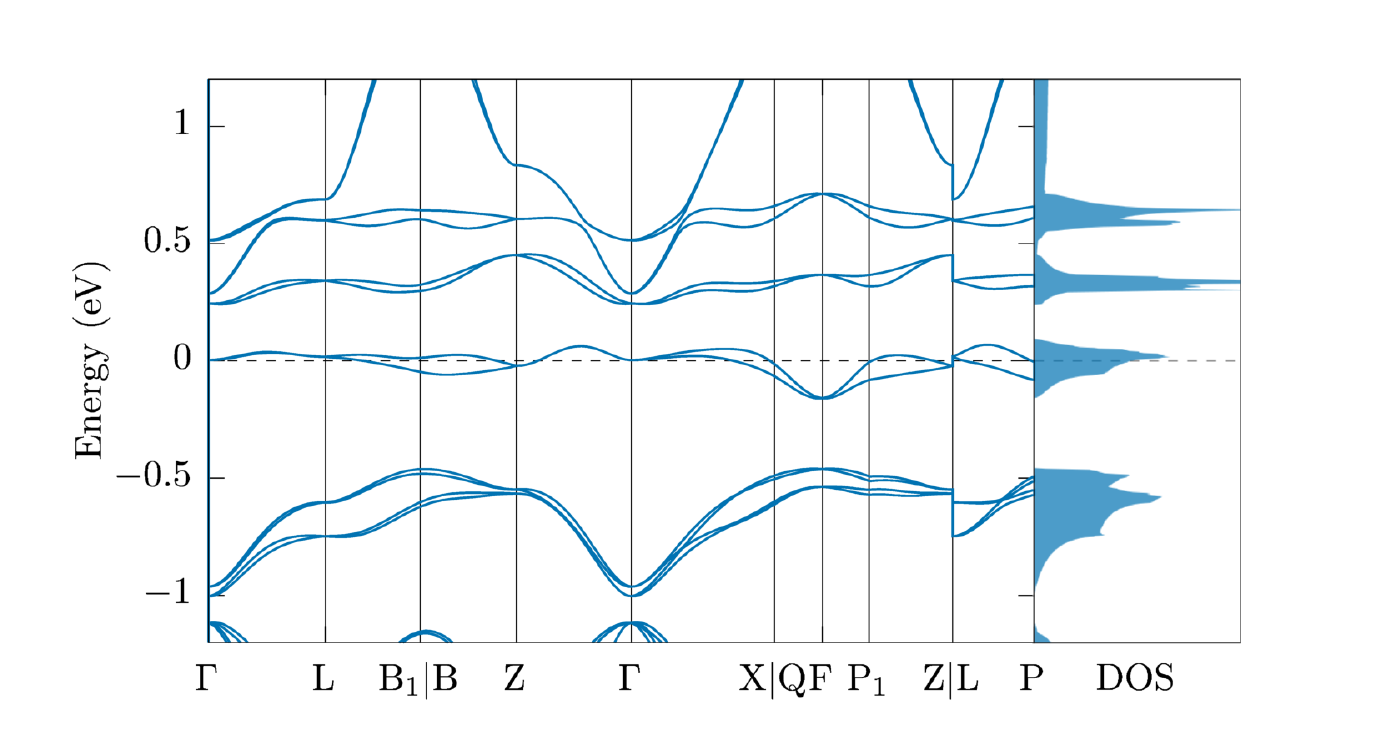}
        \label{fig:band160_soc}
      }
      \caption{ The band structure and DOS of nonmagnetic $\rh$  $\gts$ \ref{fig:band160_nosoc}  without and \ref{fig:band160_soc} with SOC.
     }
     \label{fig:band160}
    \end{figure}

    The electronic band structure and density of states (DOSs) of  $\rh$ $\gts$ are shown    in \Cref{fig:band160}.
    Both of them indicate  $\rh$ $\gts$ is metallic in our first-principles calculation, which conflicts with the experimentally observed insulating state.
  Because of the distortion of $\text{Ta}_4$ clusters, the $T_2$ representation breaks into a two-dimensional 
  representation $E$ in higher energy and a one-dimensional representation $A_1$ in lower energy,
    which coincides with the Jahn-Teller effect, as shown in the band structure.
    It is noted that the single band crossing the Fermi level is half filled and the bandwidth is  about 0.25 eV, 
  which is much narrower compared with that of 0.75 eV in $\fcc$ $\gts$~\cite{kim_spin-orbital_2014}.
    In the Hubbard model, a single band crossing the Fermi level with half filling usually implies a tendency toward an antiferromagnetic ordering state~\cite{corrlation_and_magnetism}.
    In experiments, $\gts$ has no long-range magnetic order even when the temperature is down to 1.6 K.
    For the fcc structure, there is an antiferromagnetic frustration preventing 
    the formation of  antiferromagnetic order even 
  at very low temperatures.
  This frustration originates from the fcc grids formed by $\text{Ta}_4$ clusters.
    For $\rh$ structure with an  elongation distortion,  
    Ta$_4$ clusters form a lattice of equilateral triangles respecting $C_{3v}$ symmetry. There still exists geometrical frustration within the triangle plane to prevent the formation of long-range antiferromagnetic order, while the $A$-type interlayer antiferromagnetic coupling is possible.
    
\begin{table*}
\caption{\label{tab:cif} Crystal and electronic structures for different phases of $\gts$.
The DFT results  do not include SOC. The lattice constants in parentheses are from experiments~\cite{springer_gts_crys, jakob_phd_2007}. There is no detailed structure of the $\rh$ phase reported in experiments.
}

\begin{ruledtabular}
\begin{tabular}{lcccc}
                                     &   $\fcc$         &     $\rh$          &     $\Pmn$        \\ \hline
 $a$ (\AA)                             &   10.51 (10.38)  &  7.40              &     10.50 (10.38) \\
 $b$ (\AA)                             &   10.51 (10.38)  &  7.40              &     10.50 (10.38) \\
 $c$ (\AA)                             &   10.51 (10.38)  &  18.39             &     10.53 (10.37) \\
 $\alpha$ (deg)                      &   90             &  90                &     90            \\
 $\beta$ (deg)                       &   90             &  90                &     90            \\
 $\gamma$ (deg)                      &   90             &  120               &     90            \\
Total energy (eV)/(f.u.)          &  -87.88          & -87.89             &    -87.92         \\ 
Volume ($\text{\AA}^3$)/(f.u.)    &  289.90          &  290.42            &    290.48         \\ 
    DFT  band gap (eV)               &   0              &  0                 &    0.02           \\ 
      Insulator type                 &   Mott           &  Mott              & Band Insulator    \\ 

\end{tabular}
\end{ruledtabular}
\end{table*}

\begin{figure}
      \centering
      \sidesubfloat[]{
        \includegraphics[width=0.9\linewidth]{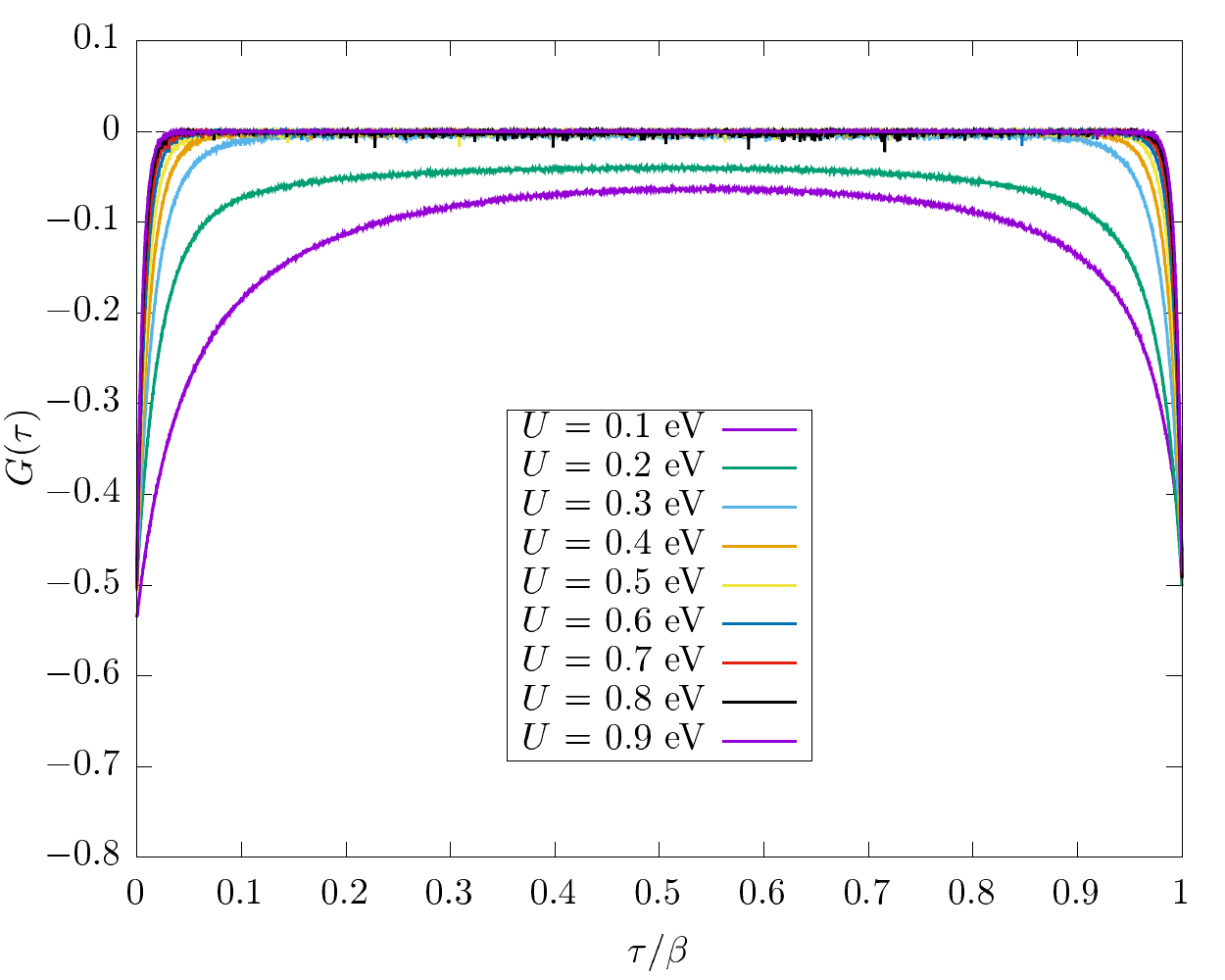}
        \label{fig:gtau}
      }

      \sidesubfloat[]{
        \includegraphics[width=0.9\linewidth]{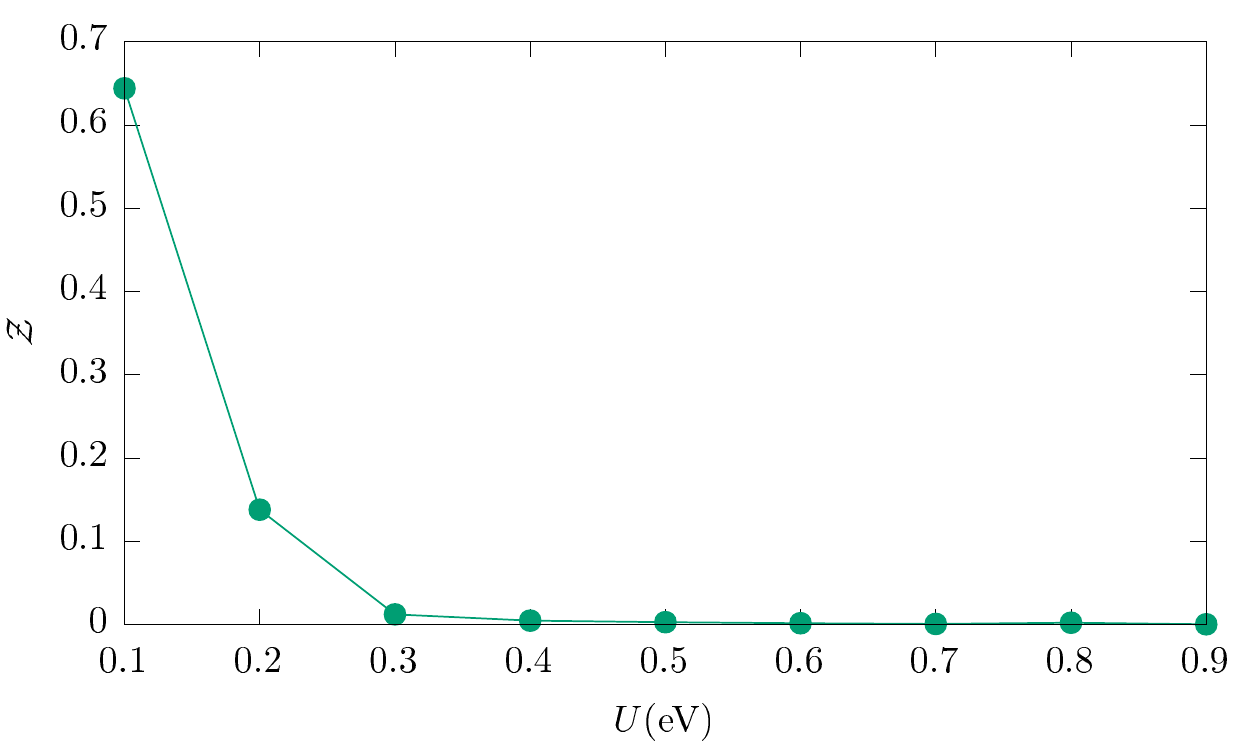}
        \label{fig:mit}
      }

      \sidesubfloat[]{
        \includegraphics[width=0.9\linewidth]{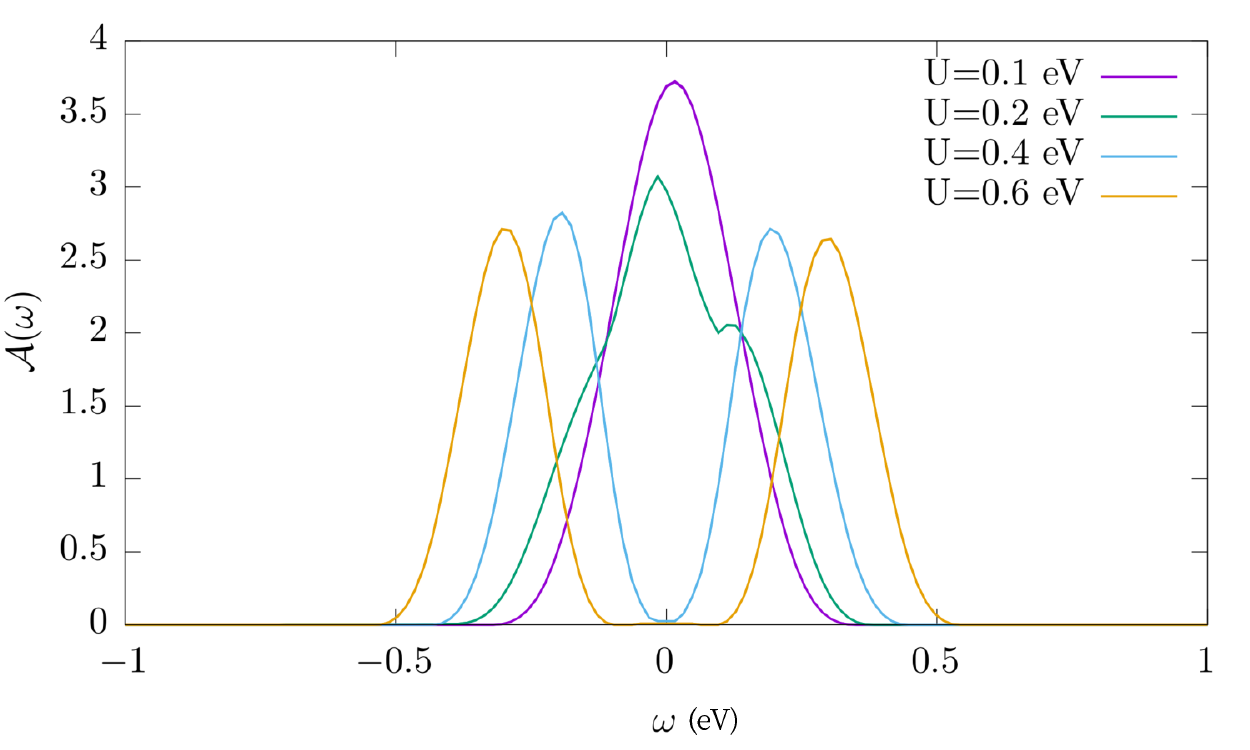}
        \label{fig:spec}
      }
      \caption{ DMFT result of $\rh$  $\gts$.
        \ref{fig:gtau}
        The impurity local image time Green function, $G(\tau)$. 
        \ref{fig:mit}
        The quasiparticle weight $\mathcal{Z}$.
        \ref{fig:spec}
        The spectrum function, $\mathcal{A}(\omega)$ of different Hubbard $U$.
      }
      \label{fig:dmft}
    \end{figure}

    On the other hand, $\rh$ $\gts$ may be a Mott insulator due to the narrow band
    formed from the molecular orbitals of Ta$_4$ clusters. To investigate this, we performed  DMFT~\cite{dmft_rev} calculation on the single-band (or single-orbital) Hubbard model using the
  \texttt{IQIST} package~\cite{iqist}, which employs the continuous time quantum Monte Carlo (CTQMC)~\cite{ctqmc_rev} impurity solver to solve the impurity imaginary time Green's function.
    Due to the absence of long-range magnetic order, we imposed the nonmagnetic phase in DMFT calculation and ignored SOC for simplicity.
    We used the DOS calculated from DFT to initialize the local Green's function. The DMFT calculation was 
  performed at temperature controlled by the parameter $\beta$ of 400 corresponding to about 29.01 K.
  The local imaginary time Green's function $G(\tau)$ [\Cref{fig:gtau}], the quasiparticle weight $\mathcal{Z}$ [\Cref{fig:mit}], and 
  the spectrum function $\mathcal{A(\omega)}$ [\Cref{fig:spec}] were calculated.
    The spectrum function is obtained from the analytic continuation  using the maximal entropy method~\cite{jarrell1996bayesian}. 

    From $\mathcal{Z}$ and the spectrum function shown in \Cref{fig:dmft}, we conclude that the metal-insulator transition (MIT) occurs when we increase the on-site 
  Coulomb interaction $U$ to about 0.3-0.4 eV, which equals 1.60$W$ with $W$ being the bandwidth of 0.25 eV.
   The on-site interaction of about 0.40 eV is still smaller than the band gap between the $n+1$ band and the $n-1$ band, where $n$ is 
  the index of the single band crossing the Fermi level.
    At the qualitative level, the single-band scheme is sufficient. 
    All the results show that the single flat band comes from the unpaired 
  $d$ electron, which is extremely localized on the $\text{Ta}_4$ cluster,
  indicating the possible Mott-insulator state of $\gts$ in the trigonal $\rh$ structure.

  \subsection{$\Pmn$ structure}
\begin{figure}
      \sidesubfloat[]{
        \includegraphics[width=0.45\linewidth]{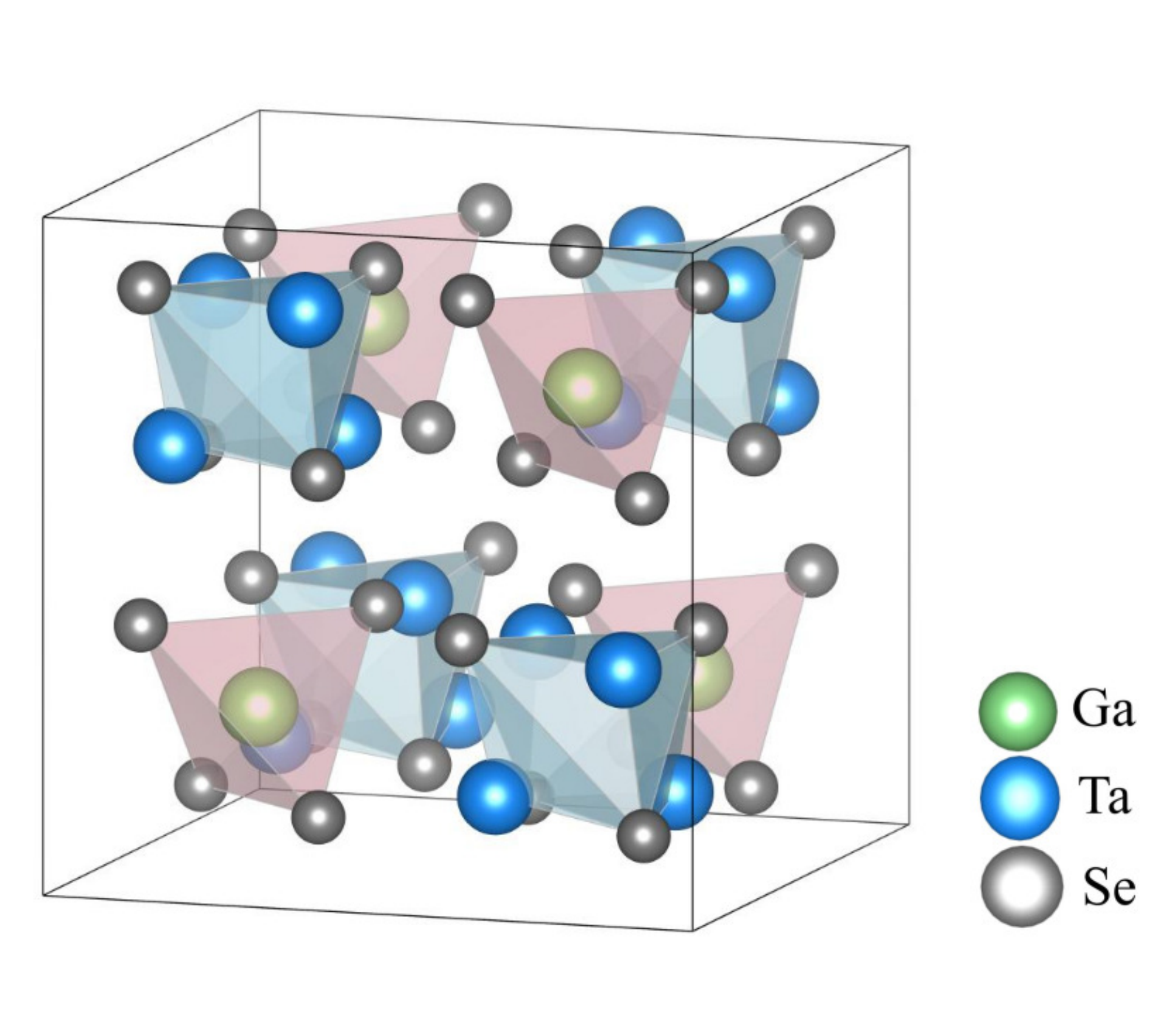}
        \label{fig:crys113}
      }
      \sidesubfloat[]{
        \includegraphics[width=0.35\linewidth]{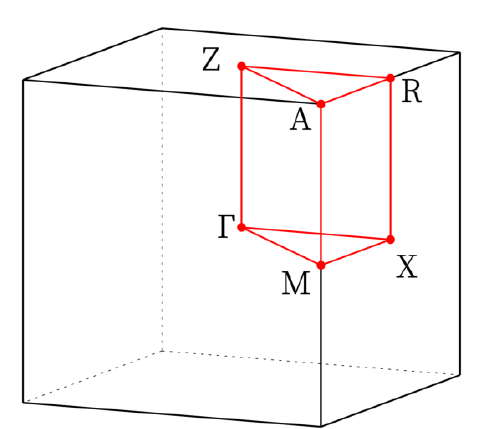}
        \label{fig:BZ113}
      }

      \sidesubfloat[]{
        \includegraphics[width=0.9\linewidth]{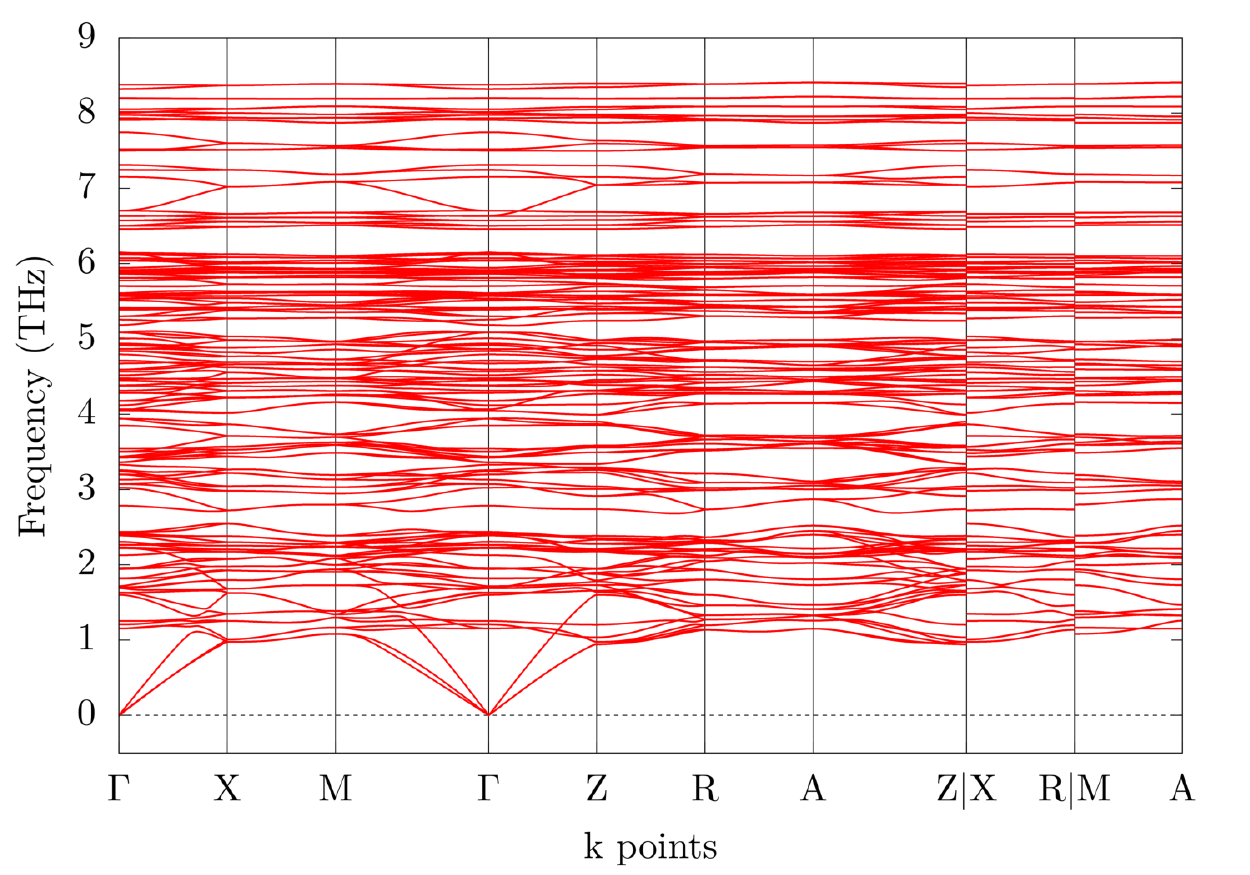}
        \label{fig:phonon113_auto}
      }
      \caption{
        \ref{fig:crys113} Crystal structure, \ref{fig:BZ113} Brillouin zone, and \ref{fig:phonon113_auto} the phonon spectra of $\Pmn$  $\gts$.
      }
      \label{fig:crys_bz113}
    \end{figure}

\begin{figure}
      \centering
      \includegraphics[width=0.7\linewidth]{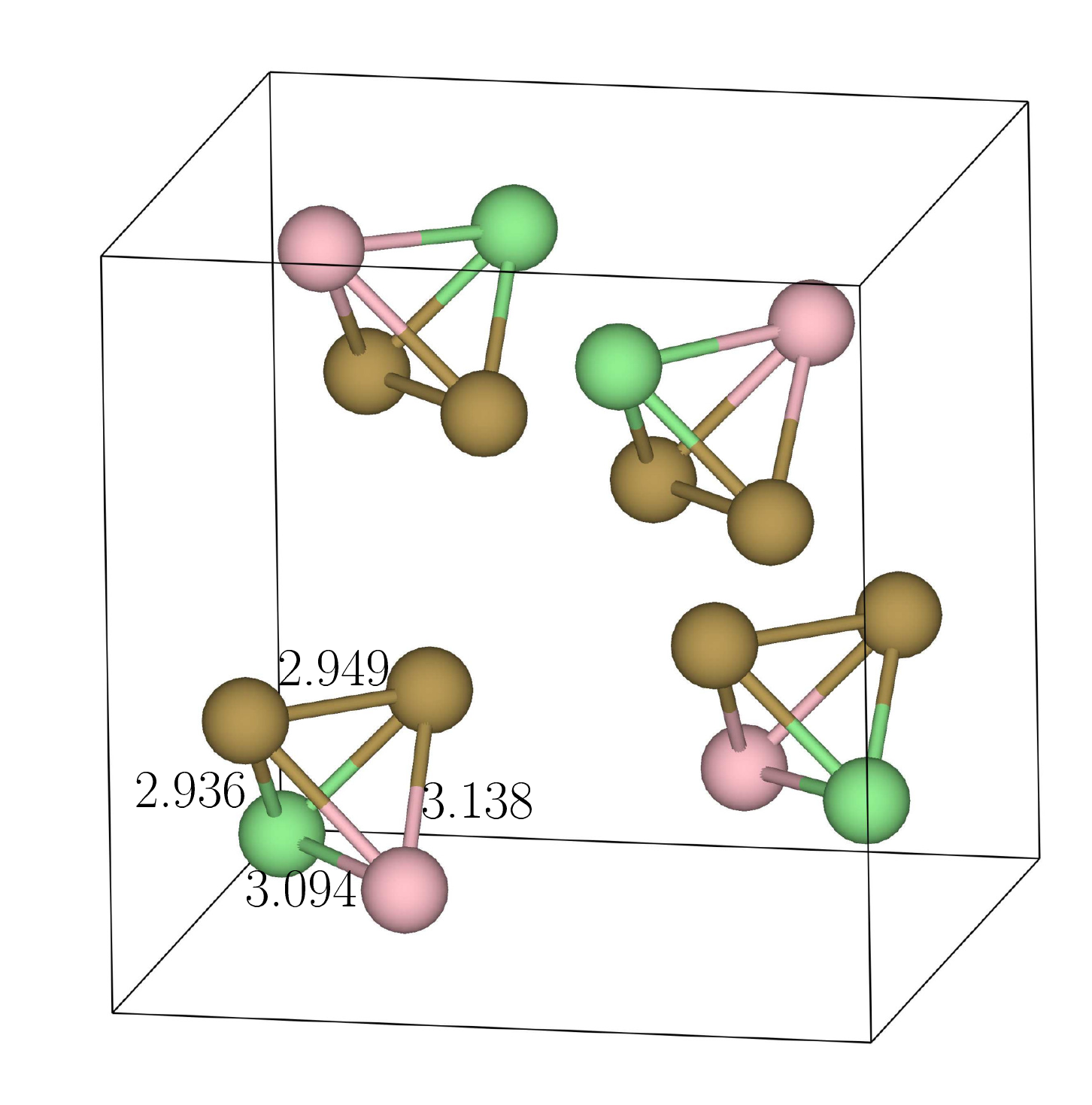}
        \caption{\label{fig:ta_cluster113}  $\text{Ta}_4$ clusters in $\Pmn$ $\gts$ after tetramerization. Different colors correspond to
      different Wyckoff positions of Ta atoms (brown, 8f; pink, 4e; green, 4e). The numbers label the bond lengths (in units \AA).
      Each $\text{Ta}_4$ cluster has a
      $C_s$ symmetry, but their ``orientation'' is  different.}
    \end{figure}

    The third crystal structure of $\gts$ studied in this work is
    the tetragonal lattice with space group $\Pmn$ (No. 113),  of which the details of crystal structure were reported in Ref. [\onlinecite{jakob_phd_2007}] based on  XRD experiments.  This phase
    was  also reported for $\gns$, where an anomalous magnetic susceptibility
    behavior appears together with the phase transition from $\fcc$ to $\Pmn$~\cite{jakob_structural_2007}.
    Considering the similarity between $\gts$ and
    $\gns$ in crystal structure and  physical properties, such as the pressure-induced superconductivity~\cite{pocha_crystal_2005, abd-elmeguid_transition_2004} and the magnetic susceptibility anomaly~\cite{jakob_structural_2007,kawamoto_frustrated_2016}, we believe
  that the $\Pmn$ structure of $\gts$ may be plausible.
  The fully relaxed crystal structure and its phonon spectra are shown in \Cref{fig:crys_bz113}.
    The symmetry of $\text{Ta}_4$ clusters is lowered to $C_{s}$. Each of them has different orientations and has four different bond lengths, as shown in \Cref{fig:ta_cluster113}. The alternative orientation with this ordering structure expands the primitive  cell, which is a kind of cooperative Jahn-Teller distortion.  

    There are four chemical formula units in a primitive cell after the cooperative
  Jahn-Teller distortion of the $\text{Ta}_4$ clusters, which can be  considered as a kind of tetramerization of the cubic phase.
    The total energy per formula unit for  $\Pmn$ $\gts$ is 0.03 eV lower than that of the $\rh$ phase.
    The phonon spectra in \Cref{fig:phonon113_auto} have no imaginary frequency,  indicating the $\Pmn$ structure is also dynamically stable. 

\begin{figure}
      \centering
      \sidesubfloat[]{
        \includegraphics[width=0.9\columnwidth]{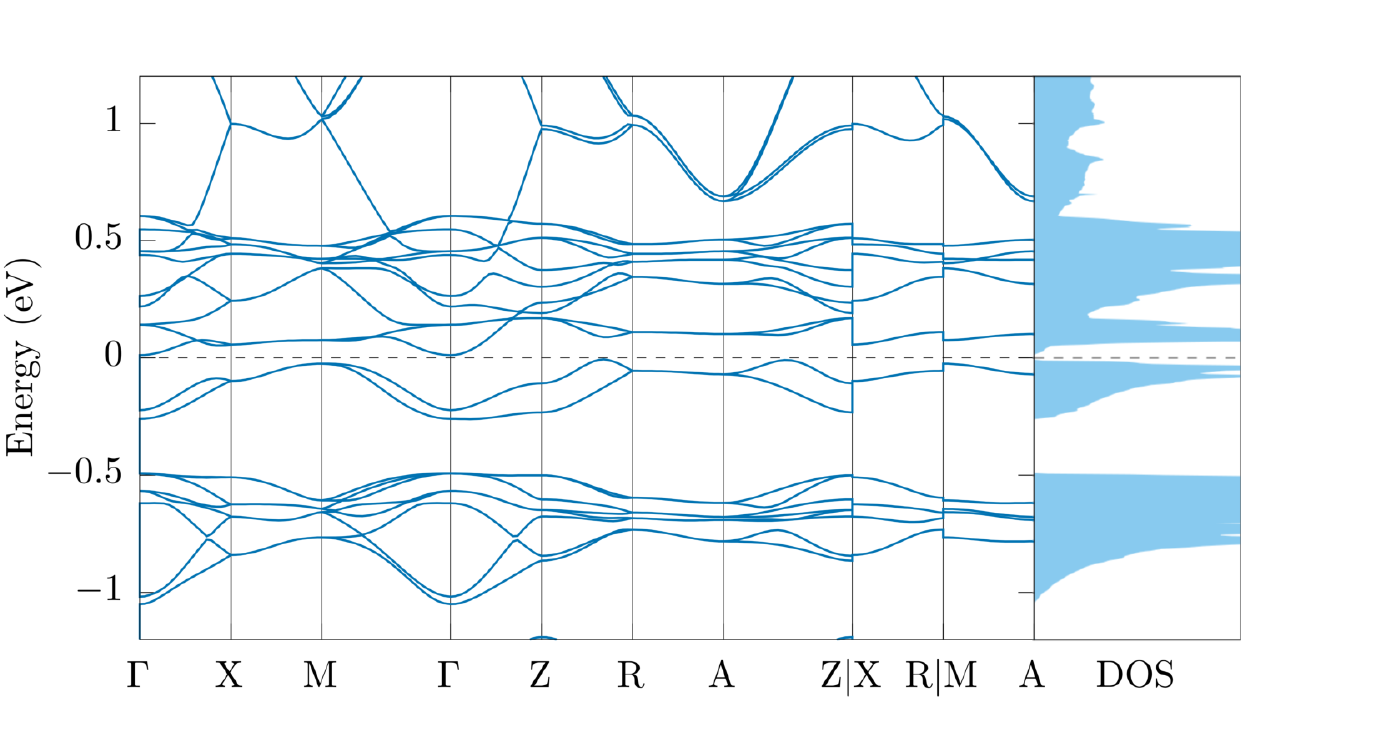}
        \label{fig:band113_nosoc}
      }
    
      \sidesubfloat[]{
        \includegraphics[width=0.9\columnwidth]{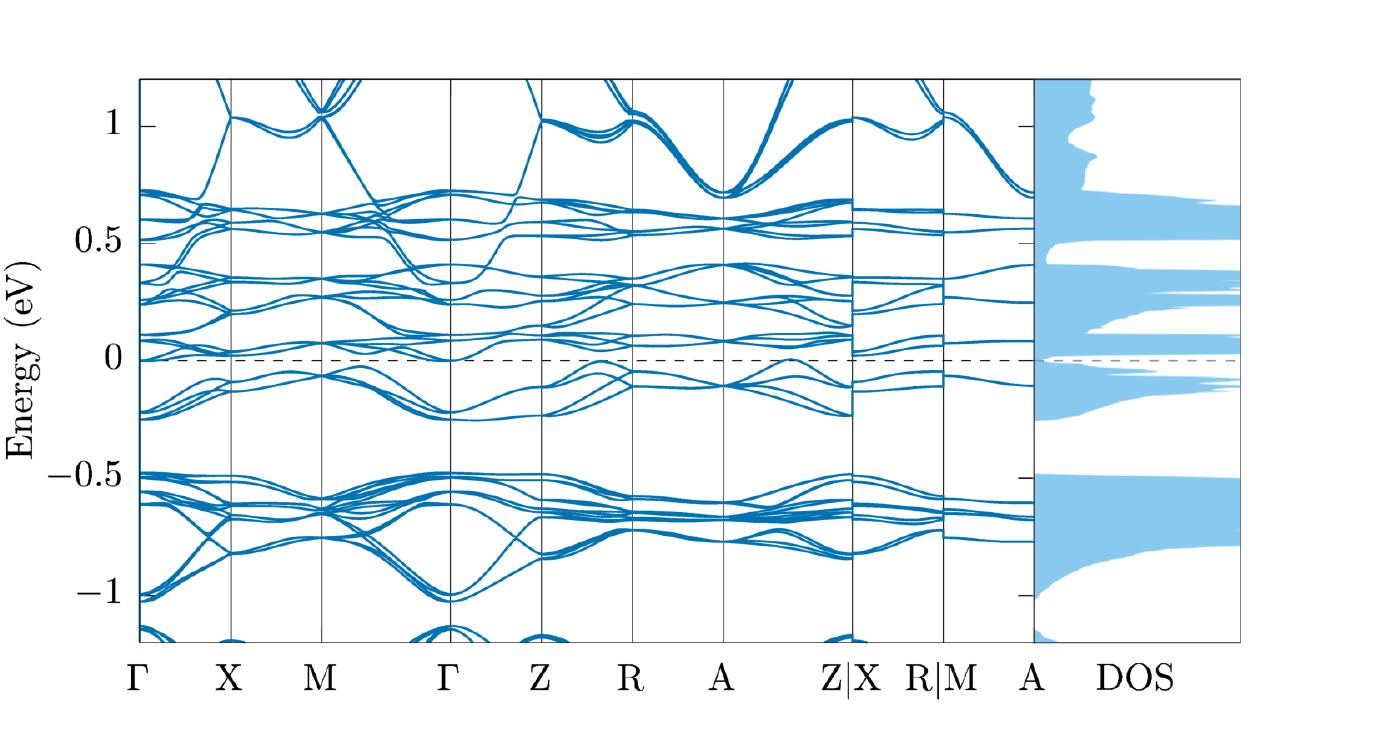}
        \label{fig:band113_soc}
      }
      \caption{Band structure and DOS of nonmagnetic $\Pmn$  $\gts$ \ref{fig:band113_nosoc} without and \ref{fig:band113_soc} with SOC.
      }
      \label{fig:band113} 
    \end{figure}

     Comparing with the former two phases, the electronic structure of $\Pmn$ $\gts$ changes qualitatively with a band gap appearing at the Fermi level as shown in \Cref{fig:band113_nosoc}.
    The new primitive cell of $\Pmn$ $\gts$ has four chemical formula units and contains an even number of valence electrons, which is different from the $\fcc$ and $\rh$ phases with an odd number of valence electrons. Furthermore, SOC brings spin splitting in the bands due to the absence of inversion symmetry as shown in \Cref{fig:band113_soc}, which keeps the band gap. 
  It is noted that the band-insulator state of $\gts$ has been ruled out based on experimental optical conductivity analysis~\cite{ta_phuoc_optical_2013}. This may be true for the $\fcc$ structure at room temperature or high pressure,
  but the low-temperature condition was not considered.
  Our calculation results indicate that the $\Pmn$  $\gts$ is the most likely  low-temperature phase and it is a band insulator instead of a Mott insulator.
  
\section{Discussion and Conclusion}
\label{sec:two}
  In \Cref{tab:cif}, we summarized the crystal structures and the corresponding electronic structures for $\fcc$, $\rh$, and $\Pmn$ phases  studied in this work.
  The presence of soft modes with imaginary frequency in the phonon spectra of the $\fcc$ phase  suggests that this phase is not dynamically stable at low temperatures and ambient pressure.

 The trigonal $\rh$ phase can be deduced from the $\fcc$ phase when the soft modes at $\Gamma$ are frozen as experimental condition changes. In fact, the $\rh$ phase was proposed as a possible high-pressure phase based on the Raman spectra in 2009~\cite{muller_phd_2007}. The high-pressure structural phase transition of $\gts$ should be noticed, especially in the research of the pressure induced superconductivity, which has drawn much attention recently~\cite{park2020pressure_topo_superconductor, jeong2020novel_superconduct_topo}. 
    As we have seen in the electronic structure of the $\rh$ phase, the slight elongation of Ta$_4$ regular tetrahedron clusters along the $\langle111\rangle$
    direction breaks the original $T_d$ point symmetry and affects the electronic structure near the Fermi level profoundly. 
    The triple-degenerate  molecular orbitals split into a double-degenerate and a nondegenerate orbital. In the DFT level, $\rh$  $\gts$ is metallic with only one flat band crossing the Fermi level. This is nearly a Hubbard model with half occupation and our DMFT results indicate that the correlation can lead to a Mott-insulator state when the on-site interaction is larger than 0.4 eV, being enough to destroy the Fermi liquid picture and open a spectral gap at the Fermi level.

    The $\Pmn$ phase proposed in Ref.~[\onlinecite{jakob_phd_2007}] may be the most possible low-temperature structure at ambient
  pressure. Recently, it was considered in the studies of lattice dynamics and electronic excitation of lacunar spinel materials ~\cite{Lattice_dynamics_2020}.
    The electronic structure of $\Pmn$ shows a  band gap at the Fermi level with no need of Hubbard $U$ or  magnetic order as the subsequence of the tetramerization of the Ta$_4$ clusters shown in \Cref{fig:ta_cluster113}. This coincides with the nonmagnetic ground state of $\gts$ from experiments~\cite{japan_gts}, indicating that the low-temperature phase of $\gts$  is not a Mott insulator, but a band insulator.
    
    We hope this work can shed light on the long-standing problem of the ambiguity in the structural phase of $\gts$, and can provide more theoretical  hints for further research in the related field.

\section{acknowledgments}
\label{sec:ack}
    This work was supported by the National Natural Science Foundation of China (NSFC) (Grants No. 11674369, No. 11925408, and No. 11921004), 
  the National Key Research and Development Program of China (Grants No. 2016YFA0300600 and No. 2018YFA0305700), 
  the Strategic Priority Research Program of Chinese Academy of Sciences (Grant No. XDB33000000), 
  the K. C. Wong Education Foundation (Grant No. GJTD-2018-01), the Beijing Natural Science Foundation (Grant No. Z180008), 
  and the Beijing Municipal Science and Technology Commission (Grant No. Z191100007219013).
  Y.D. acknowledges the support from the National Key Research and Development Program of China (Grant No. 2018YFA0305703), 
  Science Challenge Project  No TZ2016001, and the NSFC:(Grants No.  U1930401 and No. 11874075).
    Y.C. is grateful for the research computing facilities offered by ITS, HKU.

\bibliography{main}
\end{document}